\documentstyle{article}

\newcommand{\be}{\begin{equation}}
\newcommand{\ee}{\end{equation}}
\newcommand{\bea}{\begin{eqnarray}}
\newcommand{\beas}{\begin{eqnarray*}}
\newcommand{\eea}{\end{eqnarray}}
\newcommand{\eeas}{\end{eqnarray*}} 
\newcommand{\ba}{\begin{array}}
\newcommand{\ea}{\end{array}}
\newcommand{\bi}{\begin{itemize}}
\newcommand{\ei}{\end{itemize}}
\newcommand{\ben}{\begin{enumerate}}
\newcommand{\een}{\end{enumerate}}

\newcommand{\np}{Nucl. Phys.~}

\newcommand{\prd}{Phys. Rev.~}
\newcommand{\pl}{Phys. Lett.~}

\begin{document}

\renewcommand{\baselinestretch}{1.5}
\large


\title{An $[SU(3)]^4$ Supersymmetric grand unified model}
\author{Abdel P\'erez--Lorenzana$^{1,2}$ and William A. Ponce$^{3}$}

\date{\today}

\maketitle

\vspace*{-15em}\begin{flushright}{UMD-PP-99-114}\end{flushright}
\vskip10em

\begin{center}
\normalsize 1-Department of Physics, University of Maryland, College
\normalsize Park, Maryland 20742 USA. \\
\normalsize 2-Departamento de F\'\i sica, 
\normalsize Centro de Investigaci\'on y de Estudios Avanzados del I.P.N.\\
\normalsize Apdo. Post. 14-740, 07000, M\'exico, D.F., M\'exico.\\
\normalsize 3-Departamento de F\'\i sica, Universidad de Antioquia 
\normalsize A.A. 1226, Medell\'\i n, Colombia.
\end{center}

\begin{abstract}
We present a grand unified model based on the supersymmetric
$SU(3)_L\otimes SU(3)_{CL}\otimes SU(3)_{CR}\otimes SU(3)_R$ gauge group, 
which unifies in one single step the three gauge couplings of the standard
model at an scale $M\sim 10^{18}$ GeV, and spontaneously breaks down to
$SU(3)_c\otimes U(1)_{EM}$ using only fundamental representations of
$SU(3)$. In this model the proton decay is highly suppressed 
and the
doublet-triplet problem is lessened. The see-saw
mechanism for the neutrinos is readily implemented with the use of 
an extra tiny mass sterile neutral particle for each generation
which provides a natural explanation to the neutrino
puzzle.\\[1ex]
pacs: { 12.10.Kt; 12.10.Dm; 12.60.Jv;14.60.Pq}
\end{abstract}

\vskip2em

\section{Introduction}
Strings provide us with a very compelling theory, giving a consistent
framework which is finite and incorporates at the same time
both, quantum gravity and chiral supersymmetric (SUSY) gauge theories. 
When one-loop effects are included in the perturbative
heterotic string\cite{string}
they predict an unification of the gauge couplings at a scale
$M_{string}\sim 4\,\times 10^{17}$ GeV. 

On the other hand, the logarithmic running through the ``desert" of the
three gauge couplings $c_i\alpha_i^{-1}$ do merge together into a single
point, only when the SUSY partners of the standard model (SM) elementary
particles are included in the renormalization group equations (RGE), at a
mass scale $M_{susy}\sim 1$ TeV\cite{susy}. ($\alpha_i=g^2_i/4\pi, \;\;
i=1,2,3$, and $\{c_1,c_2,c_3\}=\{{3\over 5},1,1\}$ are the gauge couplings
and normalization constants of the SM factors $U(1)_Y,\;SU(2)_L$ and
$SU(3)_c$ respectively.) This amazing result, which is not upset when
higher order contributions are included in the RGE~\cite{langa}, has the
inconvenience that the unification scale, $2\times\, 10^{16}$ GeV, is a
factor of 20 smaller than the value $M_{string}$.

Several efforts to reconcile these two perturbative scales have been made
without success so far\cite{dienes}, producing always the theoretical
result $M_{string}>M_{GUT}$, where $M_{GUT}$ is the mass scale of the
grand unified theory (GUT) under consideration.

In what follows we are going to study a new SUSY-GUT which
has the property that $M_{GUT}\sim M_{string}$, without structure between
$M_{susy}\sim 1$ TeV  and $M_{GUT}$. The existence of this model can be
inferred from Fig. (5) in Ref.\cite{wp}. This note is organized in the
following way: In section 2 we introduce the new model, implement the
spontaneous symmetry breaking of the gauge group and calculate the mass
spectrum of the fermion particles. In section 3 we do the RGE analysis and
set the two different  mass scales in the model. Conclusions and remarks are
presented in the last section.

\section{The model}
We propose a SUSY-GUT based on the gauge group
$G_g\equiv [SU(3)]^4\times Z_4$ which above $M_{GUT}$ is just the
SUSY chiral-color extension\cite{2su3} of the trinification model of
Georgi-Glashow-de R\'ujula\cite{3su3}. The four $SU(3)$ factor groups are
identified as $SU(3)_L$ which contains weak $SU(2)_L$, $SU(3)_{CL}\otimes
SU(3)_{CR}$ which is the chiral color extension\cite{2su3} of $SU(3)_c$,
and $SU(3)_R$
which is the right-handed analog of $SU(3)_L$. The cyclic group $Z_4$ acting
upon the four factor groups ensures that there is only one gauge coupling
constant; more specifically, if $(L,CL,CR,R)$ is a representation under
$[SU(3)]^4$, the effect of $Z_4$ is to symmetrize it in the following way:
\[Z_4(L,CL,CR,R)=(L,CL,CR,R)\oplus (CL,CR,R,L)\oplus (CR,R,L,CL)\oplus
(R,L,CL,CR).\]
The gauge bosons of $G_g$ are assigned to the adjoint irreducible
representation (irrep) $Z_4 (8,1,1,1)$ which includes
twelve light particles (gluons, photon,
$W^\pm ,Z$), twenty superheavy, and their SUSY partners, which are all
integrally charged.

Each family of fermions is assigned to $\psi_{36}=Z_4\psi(3^*,3,1,1)$ which
under $[SU(3)_c,SU(2)_L,U(1)_Y]$ decompose as:
\beas
\psi(3^*,3,1,1)&=&(3,2,1/6)\oplus (3,1,-1/3)\\
\psi(3,1,1,3^*)&=&(1,2,1/2)\oplus 2(1,2,-1/2)\oplus (1,1,1)\oplus 2(1,1,0)\\
\psi(1,1,3^*,3)&=&(3^*,1,-2/3)\oplus 2(3^*,1,1/3)\\
\psi(1,3^*,3,1)&=&(8,1,0)\oplus (1,1,0), 
\eeas
where besides the 15 ordinary particles in each family, it contains the
right-handed neutrino field $\nu^c$ (one of the
$(1,1,0)\in\psi(3,1,1,3^*)$), one exotic down
quark, one exotic field with electric charge one, three electrically
neutral two component weyl  spinors, the electrically neutrals
spin 1/2 quaits (8,1,0), and the colorless quone 
$(1,1,0)\in\psi(1,3^*,3,1)$. For further
reference let us introduce the following convenient notation for
$\psi(3,1,1,3^*)$:
\be
\psi(3,1,1,3^*)=\left(\begin{array}{ccc}
N^0&E^-&e^-\\
E^+&N^{0c}&\nu\\
e^+&\nu^c&M^0 \end{array}\right),
\ee
where $e^{\pm}, \; \nu$, and $\nu^c$, stand for the electron, electron
neutrino and right-handed electron neutrino fields respectively.

At the unification scale, $G_g$ breaks down spontaneously to
the SUSY extension of the SM gauge group $SU(3)_c\otimes SU(2)_L\otimes
U(1)_Y=G_{SM}$ in one single step, with the particle content of the
minimal supersymmetric standard model  plus three new low energy
elementary Higgs scalar doublets, needed to produce a realistic mass
spectrum, as it is shown anon.

Indeed, the introduction of the following set of Higgs scalar
fields $Z_4\phi(3^*,3,1,1)$ and $Z_4\chi(3,3^*,3,3^*)$ with vacuum
expectation values (VEV) $\langle\phi(3^*,3,1,1)\rangle= 
\langle\phi(1,1,3^*,3)\rangle=0$,  
\[\langle\phi(1,3^*,3,1,)\rangle=\left(\begin{array}{ccc}
V&0&0\\
0&V&0\\
0&0&V \end{array}\right),\]
\[\langle\phi(3,1,1,3^*)\rangle=\left(\begin{array}{ccc}
v&0&0\\
0&v&0\\
0&0&V \end{array}\right),\]
\[\langle\wp(3,3^*,3,3^*)\rangle=\left(\begin{array}{ccc}
v&0&0\\
0&v'&0\\
0&0&V \end{array}\right),\] 
and 
\[\langle\wp(3^*,3,3^*,3,)\rangle=\left(\begin{array}{ccc}
0&0&0\\
0&0&V\\
0&v&w \end{array}\right);\]
where $\wp$ is the component of $\chi$ which points in the scalar quone
direction, $V\sim M_{GUT}$, and $v,v'$ and $w$ are related to the
electroweak breaking scale. 

The algebra shows that:
\[G_g\stackrel{V}{\longrightarrow}G_{SM}\stackrel{v}{\longrightarrow} 
SU(3)_c\otimes U(1)_{EM}.\]

With the scalars $\phi$ and $\chi$ and their VEV as introduced above, the
following trilinear invariants can be constructed:
\begin{enumerate}
\item $\psi(3,1,1,3^*)\psi(3,1,1,3^*)\langle\phi(3,1,1,3^*)\rangle$\\ 
      which gives rise to a mass term of the form:
      $v(N^0M^0+N^{0c}M^0-\nu^c\nu- 
      e^-e^+)+V(N^0N^{0c} - E^-E^+)+h.c.$
\item $\psi(3^*,3,1,1)\psi(1,1,3^*,3)\langle\chi(3,3^*,3,3^*)\rangle$\\
      which gives rise to masses of order $v,\;v'$, and $V$ to the up,
      down and exotic down quarks respectively.
\item $\psi(1,3^*,3,1)\psi(1,3^*,3,1)\langle\phi(1,3^*,3,1)\rangle$\\
      which gives rise to masses of order $V$ to the eight spin 1/2 quaits
      and to the quone.
\item $\psi(3,1,1,3^*)\psi(1,3^*,3,1)\langle\chi(3^*,3,3^*,3)\rangle$\\
      which gives rise to a mass term of the form
      $\sqrt{3}D^0(v\nu+V\nu^c+w M)+h.c.$, where $D^0$ is the spin 1/2
      quone $(1,1,0)\in\psi(1,3^*,3,1)$.
\end{enumerate}
From the former results, the six electrically neutral spin 1/2 color
singlets in one generation mix in the following way (in the basis given by 
$\{D^0,\nu^c,\nu,N^0,N^{0c},M^0\}$):
\be
\left(\begin{array}{cccccc}
2V&\sqrt{3}V&\sqrt{3}v&0&0&w\\
\sqrt{3}V&0&-v&0&0&0\\
\sqrt{3}v&-v&0&0&0&0\\
0&0&0&0&V&v\\
0&0&0&V&0&v\\
w&0&0&v&v&0 \end{array}\right),
\ee
which for the particular case $w=0$ (which does not alter the symmetry
breaking pattern) has four eigenvalues of order $V$ and two seesaw
eigenvalues, $-2v^2/V$ and $8v^2/3V$, corresponding to the mixing of $M$
with $N^0$ and $N^{0c}$, and of $\nu$ with $\nu^c$ and
$D^0$ respectively (when $w\leq v$, the eigenvalues are of the same order,
but a more general mixing occurs).

Notice that the number of low energy $(\sim v(v'))$ Higgs doublet
scalar fields introduced in the former expressions is five, independent of
the value for $w$ which is the VEV of a scalar field which is a singlet
under the SM quantum numbers.

\section{The Mass scales}
The two loop RGE predictions for the gauge couplings in the SUSY standard
model (ignoring Yukawa couplings) can be written as: 
\be
\alpha_i^{-1}(m_Z)=\frac{\alpha^{-1}}{c_i}
-b^0_i\ln\left({M\over {m_Z}}\right)+\sum_{j=1}^3 
\frac{b_{ij}^1}{b_j^0}\ln\left({c_j\alpha\over
{\alpha_j(m_Z)}}  \right)+ \Delta_i
\label{tlrge}
\ee
where $M$ is the GUT scale, $\alpha=g^2/4\pi$ is the gauge coupling for
$G_g$, $\{c_1,c_2,c_3\}=\{{3\over 5},1,{1\over 2}\}$,
and $b^0_i,\; b_{ij}^1,\;i,j=1,2,3$ are the one loop and two loops SUSY
beta functions respectively. In the former expression we have lumped
together into $\Delta_i \; (i=1,2,3)$ the $\overline{MS}$ to
$\overline{DR}$\cite{akt} renormalization scheme conversion factor
$(C_2(G_i)/12\pi)$, the SUSY thresholds, and other effects as for example
possible (small) contributions from extra dimensions, contributions of
possible nonrenormalizable operators, etc.

Starting our analysis with the one loop calculations we set
$\Delta_i=b_{ij}=0$, and use the one loop SUSY beta functions\cite{bss}:
\be
2\pi\left(\begin{array}{c}b_1^0\\b_2^0\\b_3^0 \end{array}\right)=
\left(\begin{array}{c} 0\\6\\9\end{array}\right)-
\left(\begin{array}{c}10/3\\2\\2\end{array}\right)F-
\left(\begin{array}{c}1/2\\1/2\\0\end{array}\right)H,
\ee
where $F=3$ is the number of SUSY families and $H=5$ is the number of
light $SU(2)_L$ scalar doublets present in the model.

Our approach is the known one\cite{langa} of using the experimental 
imputs\cite{pdg} for $\alpha_1^{-1}(m_Z)=98.330\pm 0.091$ and
$\alpha_2^{-1}=29.517\pm 0.043$ in Eqs.(\ref{tlrge}) for $i=1,2$
in order to calculate values for $M$ and $\alpha$, and then use
those results in the other Eq.(\ref{tlrge}) ($i=3$) in order to predict a
value for $\alpha_3(m_Z)$. When the algebra is done we get $M\sim
1.5\times 10^{18}$ GeV and $\alpha^{-1}=14.86$ which in turn implies
$\alpha_3(m_Z)=0.083$ which is about $30\%$ off the experimental
value\cite{pdg} $\alpha_3^{exp}(m_Z)=0.119\pm 0.017$. 

Next let us look for solution to Eqs.(\ref{tlrge}) including the second
order effects. We then use $\Delta_i=\delta_i/12\pi$, ($\delta_i=0,2,3$ for
$i=1,2,3$ respectively, the $\overline{MS}$ to $\overline{DR}$ renormalization
scheme conversion factor),  the two loop beta functions\cite{bss}:
\be
8\pi^2\left(\begin{array}{ccc}
b_{11}^1 & b_{12}^1 & b_{13}^1 \\ 
b_{21}^1 & b_{22}^1 & b_{23}^1 \\ 
b_{31}^1 & b_{32}^1 & b_{33}^1 \end{array}\right)=
\left(\begin{array}{ccc}
-{190\over 27}F-{1\over 2}H & -2F-{3\over 2}H & -{88\over 9}F \\
-{2\over 3}F-{1\over 2}H & 24-14F-{7\over 2}H & -8F\\
-{11\over 9}F & -3F & 54-{68\over 3}F
\end{array}\right), \label{ttlrge}
\ee
($F=3$ and $H=5$ as before), and introduce the SUSY partners of the
known particles in the SM at the weak scale $m_Z$ 
in order to take into account
low energy threshold effects\cite{langa}.  When the algebra is done 
we get $M\sim 3 \,\times\, 10^{18}$ GeV,
$\alpha^{-1} = 16.76$ and $\alpha_3(m_Z)=0.128$, this last value within
the experimental limits allowed by $\alpha_3^{exp}(m_Z)$. 
[The solution quoted has a strong dependence on the
$H$ value; as a matter of fact, $H=4$ produces a very small value for
$\alpha_3(m_Z)$].

This amazing result suffers from the flaw that the GUT scale predicted is
almost one order of magnitude greater than $M_{string}$, where gravity becomes at
least as important as the other interactions and can not be ignored. Now,
if we claim that $M_{string}$ is not $4\times 10^{17}$ GeV, but a
smaller value (something in between 1 TeV and $10^{11}$ GeV)
coming from the nonperturbative effects of the string\cite{qui}, 
then the entire idea
of a GUT must be reconsidered. A more reasonable approach is to assume that
even the non perturbative effects in the string are at most of the same
order of the perturbative ones (which are small at this scale as we will 
see next). If this is the case then we may argue that
other effects as for example contributions from Kaluza-Klein (KK) modes,
or extra dimensions, are tractable and may slightly change the
perturbative GUT scale and the value for $\alpha_3(m_Z)$. 
Lets us see this with the following example: if we use for
$\Delta_i$ the expression\cite{mab} 
\be
\Delta_i=\frac{\delta_i}{12\pi}+\tilde{b}_i
\left\{\frac{1}{2\pi}\left[\left(\frac{M}{M_{string}}\right)-1
-\ln\left(\frac{M}{M_{string}}\right)\right]\right\}
 \ee 
where $\tilde{b}_i$ are the beta functions for the KK modes, and assume that
the only KK modes present are the gauge bosons and an $SU(2)_L$ doublet 
of scalar fields,
then we get for solution to the new set of equations
$M\sim 1.28 \times 10^{18}$ GeV and
$\alpha_3(m_Z)=0.114$; 
so the net effect of this KK modes is to lower a
little the GUT scale and to bring $\alpha_3(m_Z)$ closer to its
experimental value. Other KK modes may do the opposite, but the net effect will
be small because $M\sim M_{string}$.

\section{Concluding remarks}

In this note we have presented various aspects of a new SUSY-GUT which
unifies, in one single step, the three gauge couplings of the SM at a 
mass scale $10^{19}$ GeV $>M_{GUT}\geq M_{string}$. We believe this model
opens a door in the so called string-GUT problem\cite{iba}, due to the
fact that it uses only fundamental irreps (and their conjugates) for
scalar and spinor fields. In addition, when we compare our normalization
coefficients $c_i$ with the Kac-Moody levels of the four dimensional
string, we have that $\kappa_i=c_i^{-1}$, which for $c_2=1$ and
$c_3={1\over 2}$ implies that only level one and two could be needed when the
ten dimensional SUSY-string is compactified to four dimensions. From the
literature\cite{alda} we know that it is simple to compactify at
levels $\kappa =1,2$ and produce at the same time massless states in the
fundamental irreps of the gauge group. 

Proton decay is highly suppressed
in the context of this model: the gauge
bosons are integrally charged and can not mediate proton decay, and there
are no Higgs scalars multiplets of the form $Z_4\phi(3^*,1,3,1)$ which
are the only ones which couple to both, quarks and leptons at tree level.

By imposing the validity of the extended survival hypothesis\cite{esh}, 
the doublet-triplet Higgs splitting problem, present in GUT $SU(5)$
and its extensions, is lessened in our model, since the representations
containing
$SU(3)_c$ Higgs field triplets which are $SU(2)_{L(R)}$ doublets do not 
develop VEV at all.  also the chiral color Higgs fields are either quaits
or quones of $SU(3)_c$, with only the quones developing VEV and existing
at the low energy scale. 

It is worth  mentioning the peculiar way in which the seesaw mechanism
for the neutrinos is implemented in the context of the model, via 
mixings with the right-handed neutrino field $\nu^{c}$ (coupled with
$SU(2)_R$ scalar singlets instead of triplets as it is usually done), and
with the peculiar sterile quone $D^0$ which is a SM singlet. Also, 
besides the usual tiny massive neutrinos, there is an extra light
particle  in each family, it is the sterile $M^0$ which mixes with
$\nu$ when $w\neq 0$. Those particles which may contribute to the
dark matter of the universe, but very little to nucleosynthesis\cite{foot},
are 
the right ingredients needed to explain the neutrino
puzzle\cite{puzzle}; that is, to explain the neutrino oscillations in the
sun, in the atmosphere, and at the LSND\cite{lsnd} experiment in los
Alamos\cite{cha}.

The fact that $H=5$ is used, instead of other value, is not arbitrary.
Indeed, the suppression of any Higgs field $SU(2)_L$ doublet with VEV of
order $v(v')$ in our analysis, will imply either a zero mass for a known
particle (up or down quark and electron), or a failure in the
implementation of the see-saw mechanism. So, to take $H\geq 5$ is
compulsory, but $H>5$ is redundant.

What is the advantage of moving from $[SU(3)]^3$ to $[SU(3)]^4$? As it can be
seen from the second paper in Ref.\cite{3su3}, it is very difficult to get a
decent mass spectrum for the known particles in the trinification model (some
particular assumptions on the radiative corrections of the model must be made).
On the contrary, the mass spectrum in our model comes easely at tree level, 
for a reduce set of scalar fields.

Why SUSY $[SU(3)]^4$ rather than the  non-SUSY version? 
Because the non-SUSY version of
$[SU(3)]^4$ does not unify the gauge coupling constants, unless a very large
amount of Higgs field doublets is introduced ($H=27$).

Finally let us mention that the VEV structure of the Higgs scalars used is the
minimum compatible with a consistent mass spectrum. To increase the number of
possible VEV will produce tiny see-saw masses of order $v^2/V$ to the electron
or to the bottom quark. To reduce the number of possible VEV will produce zero
masses to some known particles. It will be very nice if the pattern of VEV we
used can be obtained from the minimum of the scalar potential, but such analysis
is beyond the scope of the present work.

\section*{Acknowledgments}
This work was partially supported by CONACyT, M\'exico and Colciencias,
Colombia.  We thank Z. Chacko for reading and commenting the original
manuscript. W.A.P. thanks  R.N. Mohapatra, and the Physics Department of the
University of Maryland at College Park for hospitality during the completion of
this work.


\begin{thebibliography}{99}

\bibitem{string}
P.Ginsparg, \pl {\bf B197}, 139 (1987); V.S.Kaplunovsky, \np {\bf B307}, 145
(1988); {\it ibid.} {\bf B382}, 436 (1992).

\bibitem{susy}
U.Amaldi, W. de Boer, and H.Furstenau, \pl {\bf B260}, 447 (1991); P.Langacker
and M.Luo, \prd {\bf D44}, 817 (1991); U.Amaldi {\it et al}, \pl {\bf B281},
374 (1992). 

\bibitem{langa}
P.Langacker and N.Polonsky, \prd {\bf D47}, 4028 (1993).

\bibitem{dienes}
For a review see: K.R.Dienes, Phys. Repp. {\bf 287}, 447 (1997).

\bibitem{wp}
A. P\'erez-Lorenzana, A. Zepeda, and W.A. Ponce, Mod. Phys. Lett. {\bf A13
No. 26}, 2153 (1998).

\bibitem{2su3}
J.C.Pati and A.Salam, \np {\bf B150}, 76 (1979); P.H.Frampton and
S.Glashow, \pl {\bf 190B}, 157 (1987).

\bibitem{3su3}
A.de R\'ujula, H.Georgi, and S.Glashow, in {\it fifth workshop on Grand
Unification}, edited by K.Kang, H.Fried, and P.Frampton (World Scientific,
Singapore, 1984), P. 88; K.S.Babu, X.-G. He, and S. Pakvasa, Phys. Rev.
{\bf D33}, 736 (1986).

\bibitem{akt}
D.M.Capper, D.R.T.Jones, and P.Van Nieuwenhuizen, \np {\bf B167}, 479
(1980); I.Antoniadis, C.Kounnas, and K.Tamvakis, \pl {\bf 119B}, 377
(1982).

\bibitem{bss}
C.Kounnas, A.B.Lahanas, D.V.Nanopoulos, and M.Quir\'os, \np {\bf B236},
438 (1984).

\bibitem{pdg}
Particle data group: C.Caso {\it et al}, The Europhysics Journal {\bf C3
Nos. 1-4}, p. 69 (1998).

\bibitem{qui}
 I. Antoniadis, Phys. Lett. {\bf B246}, 377 (1990);  
 I. Antoniadis, K. Benakli and M. Quir\'os, Phys. Lett. {\bf B331}, 313 (1994);
N. Arkani-Hamed, S. Dimopoulos and  G. Dvali, \pl {\bf B429}, 263 (1998).

\bibitem{mab}
T.R. Taylor and G. Veneziano, \pl {\bf B212}, 147 (1988).
For a current discussion of this topic see for instance:
A.P\'erez-Lorenzana and R.N. Mohapatra: ``Effects of extra dimensions in
gauge coupling unification" hep-ph/9904504; and references therein. 

\bibitem{iba}
L.E.Iba\~nez, \pl {\bf B318}, 73 (1993).

\bibitem{alda}
G.Aldazabal, A.Font, L.E.Iba\~nez, and A.M.Uranga, \np {\bf B452}, 3
(1995); {\bf B465}, 34 (1996).

\bibitem{esh}
The extended survival hypothesis reads: ``Higgses acquire the maximum mass
compatible with the pattern of Symmetry breaking". For a detailed
explanation see: F. del Aguila, and L. Iba\~nez, Nucl. Phys. {\bf B177},
60 (1981). 


\bibitem{foot}
Our model, within the standard assumptions that go into the discussion of big
bang nucleosynthesis would imply the existence of three extra neutrinos.
However, in models with sterile neutrinos, possibility of large lepton asymmetry
at the big bang era has been discussed in: R. Foot and R. Volkas, Phys. Rev.
{\bf D56}, 6653 (1997); N.F. Bell, R. Foot and R. Volkas, Phys. Rev. {\bf D58},
105010 (1998).

\bibitem{puzzle}
B.Kayser: ``Neutrino mass. Where do we stand and where are we going"
Hep-ph/9810513, Oct. 1998, and 29$^{th}$ International Conference on high
energy physics, Vancouver, July 1998.

\bibitem{lsnd}
C.Athanassopoulos {\it et al}, Phys. Rev. Lett. {\bf 77}, 3082 (1996);
{\bf 81}, 1774 (1998).

\bibitem{cha}
Z. Chacko and R.N. Mohapatra in ``Sterile neutrinos in $E_6$ and a natural
understanding of vacuum oscillation solution to the solar neutrino puzzle.''
hep-ph/9905388, exploit this idea in the context of an $E_6$ SUSY-GUT.


\end{thebibliography}
\end{document}